# PILAR: Personalizing Augmented Reality Interactions with LLM-based Human-Centric and Trustworthy Explanations for Daily Use Cases

Ripan Kumar Kundu *
University of Missouri-Columbia

Istiak Ahmed †
University of Missouri-Columbia

Khaza Anuarul Hoque ‡
University of Missouri-Columbia

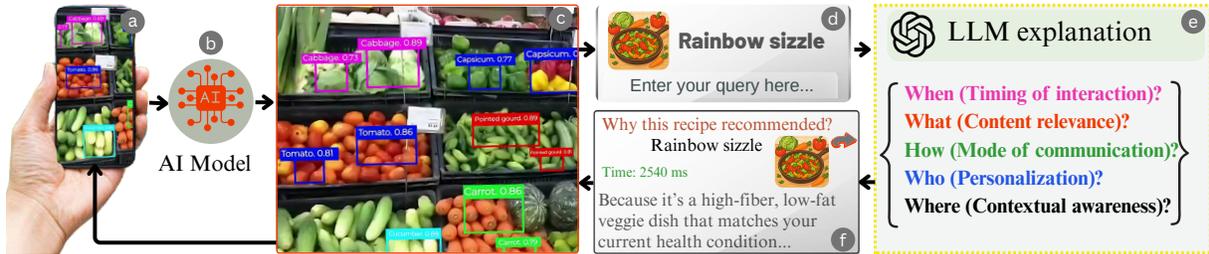

Figure 1: An overview of the *PILAR* framework. (Left) **(a)** The user points a smartphone at fresh produce for recipe recommendations, **(b–c)** An on-device object detector identifies ingredients with bounding boxes and labels in real-time, **(d)** Recipes are ranked based on the user's dietary profile, **(e)** An LLM module generates explanations across all five XAI dimensions (When, What, etc.,), and **(f)** The user asks, *"Why this recipe recommended?"* and receives a personalized, contextual explanation of the recommended recipes.

## ABSTRACT

Artificial intelligence (AI)-driven augmented reality (AR) systems are becoming increasingly integrated into daily life, and with this growth comes a greater need for explainability in real-time user interactions. Traditional explainable AI (XAI) methods, which often rely on feature-based or example-based explanations, struggle to deliver dynamic, context-specific, personalized, and human-centric insights for everyday AR users. These methods typically address separate explainability dimensions (e.g., when, what, how) with different explanation techniques, resulting in unrealistic and fragmented experiences for seamless AR interactions. To address this challenge, we propose *PILAR*, a novel framework that leverages a pre-trained large language model (LLM) to generate context-aware, personalized explanations, offering a more intuitive and trustworthy experience in real-time AI-powered AR systems. Unlike traditional methods, which rely on multiple techniques for different aspects of explanation, PILAR employs a unified LLM-based approach that dynamically adapts explanations to the user's needs, fostering greater trust and engagement. We implement the PILAR concept in a real-world AR application (e.g., personalized recipe recommendations), an open-source prototype that integrates real-time object detection, recipe recommendation, and LLM-based personalized explanations of the recommended recipes based on users' dietary preferences. We evaluate the effectiveness of PILAR through a user study with 16 participants performing AR-based recipe recommendation tasks, comparing an LLM-based explanation interface to a traditional template-based one. Results show that the LLM-based interface significantly enhances user performance and experience, with participants completing tasks 40% faster and reporting greater satisfaction, ease of use, and perceived transparency. These findings demonstrate that PILAR's LLM-driven explanations more effectively support diverse user needs in real-world AR applications, fostering seamless, personalized, and trustworthy AI experiences.

---

*e-mail: rkundu@missouri.edu
†e-mail: ia5qq@missouri.edu
‡e-mail: hoquek@missouri.edu

## 1 INTRODUCTION

Recent advances in artificial intelligence (AI) have significantly enhanced interactive augmented reality (AR) systems, reshaping how we navigate, shop, communicate, and engage with the real world [8, 13, 16, 17, 19, 32]. As AR applications increasingly support daily activities (e.g., navigation, cooking assistance, etc.,), the role of AI has become essential. However, traditional AI models act as black boxes, obscuring their decision-making processes [12]. In AR scenarios where users rely on real-time guidance and contextual recommendations, this opacity can lead to inaccurate recommendations, compromised safety, and a breakdown of user trust [32]. For instance, in an AR-based intelligent cooking assistant, unclear reasoning behind recommended recipes based on visible ingredients can lead to hesitation and reduced trust, thereby hindering the system's effectiveness. Moreover, the growing reliance on black-box AI models has raised concerns about AI misuse and eroding user trust, highlighting the need for more transparent and interpretable methods. This has fueled increasing interest in explainable AI (XAI) for AR, where end-users directly interact with AI-generated outcomes.

XAI addresses these issues by making AI decisions transparent, which is crucial in AR systems involving real-time interaction, as understanding the rationale behind recommendations is essential for user trust and experience [19, 24, 26, 32]. For instance, when an AR assistant recommends a restaurant or recipe, providing an explanation aligned with the user's preferences can foster greater comfort and confidence [26]. To support this, researchers have proposed various XAI design frameworks focused on when, what, and how to explain [13, 17, 24, 32]. However, other dimensions, such as who the explanation is for and where it is delivered, remain underexplored yet crucial for effective XAI design in AR. Moreover, studies show that many current XAI methods are unintuitive and cognitively demanding, reducing task efficiency, user performance, and satisfaction. Additionally, most prior work also overlooks how users engage with explanations and their personal preferences, which can change dynamically. Indeed, these systems often require users to select which explanations to view and lack support for natural follow-up questions. As a result, they fall short of delivering personalized, human-centric explanations that adapt to the evolving needs and context of AR environments. Recent studies have explored the use of large language models (LLMs) in AR for tasks such as context-aware voice assistants [19], assistive systems for daily activity logging [17], in-context writing tools [8], and procedural guidance [28]. However, these works focus on using LLMs for system development rather than as explanation tools. Thus, a pressing question arises: *How can we leverage LLMs as explainers to build a trustworthy, human-understandable AR ecosystem for everyday use?* Addressing this requires a holistic approach that reflects the unique characteristics of AR (e.g., rich sensory input and dynamic context). Our vision is an AR ecosystem in which users trust and understand AI-augmented experiences, making the technology a dependable daily companion. Bridging this gap could transform AR into a truly trusted and intelligible companion in daily life, highlighting a critical need for research on LLM-based explanation in AR.

To address these gaps, we propose *PILAR*, a novel framework that leverages pre-trained LLMs to generate personalized, human-centric explanations, enabling more intuitive and trustworthy interactions in real-time AI-powered AR systems (See Figure 1). Unlike traditional methods that rely on multiple explanation techniques for different aspects, PILAR integrates a pre-trained LLM model as a post-hoc explanation technique to deliver context-aware, dynamic reasoning tailored to user needs, fostering trust and engagement. We demonstrate PILAR through an intelligent user interface (UI) for a real-world AR application in personalized recipe recommendations, which integrates real-time object detection, recipe suggestions, and LLM-based explanations. Our contributions are threefold:

- We develop *PILAR*, a novel LLM-based explanation framework covering all key explainability dimensions (when, what, how, who, and where). PILAR provides context-aware, personalized, human-centric insights in real time through an adaptive AR interface, delivering intuitive and relevant information tailored to each user's interactions and preferences.

- We implement the PILAR concept in a smartphone-based AR application for personalized recipe recommendation. Our prototype runs on a Qualcomm Snapdragon 855-powered Samsung Galaxy S10 smartphone, a choice motivated by the fact that many SOTA AR devices (e.g., Meta Quest Pro) also utilize Snapdragon chipsets, and we have open-sourced the implementation[1].

- We evaluate PILAR through a user study with 16 participants, comparing its LLM-based explanation UI to a traditional template-based UI in AR recipe recommendation tasks. Results show that PILAR outperforms the traditional approach, with participants completing tasks in an average of 40% shorter time and reporting higher levels of satisfaction, ease of use, and perceived responsiveness, as it provides coherent and personalized insights during recipe recommendation tasks.

## 2 RELATED WORK

This section reviews prior work on XAI and LLM integration in AR systems. Table 1 provides a summary comparing our proposed PILAR framework with SOTA methods in terms of XAI capabilities and LLM integration. The details of these are described below.

**Explainable AI in AR:** AR systems have rapidly advanced, blending real and virtual environments through modern hardware and frameworks that enable real-time, context-aware information delivery [9, 28, 33]. As AI increasingly drives decision-making in AR, users may become confused by opaque or unexpected outcomes [32]. Thus, XAI is essential for making AI behavior interpretable, reducing user confusion, and fostering trust. Recent research has begun integrating XAI into AR to enhance transparency [23]. For instance, Wintersberger et al. [31] showed that AR-based traffic information during driving can enhance user trust through timely explanations. While authors in [24] developed an XAI-based mobile AR tool for smart home automation. However, their rule-based recommendations lacked human-centric reasoning. Moreover, without well-defined explanation tasks, it remains difficult to compare XAI-AR systems or detect misinterpretations in dynamic, context-aware environments. Therefore, very recently, Xu et.al. [32] proposed the XAIR framework, which outlines when to deliver explanations (e.g., at key decision points), what to explain (e.g., input features), and how to present them (e.g., visually or auditorily in AR). However, XAIR uses a predefined template-based method for generating different explanations, lacks support for personalization (who) and spatial context (where), and does not use LLMs, limiting its ability to generate dynamic, human-centric personalized explanations in real-time AR environments.

---
[1]https://github.com/dependable-cps/PILAR-X

Table 1: Comparison of representative SOTA methods with the proposed *PILAR* framework in terms of XAI and LLM integration.

| Method | XAI | LLMs | Method | XAI | LLMs |
|---|---|---|---|---|---|
| XAIR [32] | ✓ | | PANDALens [8] | | ✓ |
| Explainable-Interfaces [34] | ✓ | | TOM [17] | | ✓ |
| Explainable Automation [25] | ✓ | | GazePointAR [19] | | ✓ |
| XARSS [35] | ✓ | | G-VOILA [30] | | ✓ |
| Wearable Reasoner [10] | ✓ | | XR-Objects [13] | | ✓ |
| PilotAR [16] | | | XaiR [28] | | ✓ |
| Goldilocks Zoning [15] | ✓ | | LLMR [11] | | ✓ |
| *Proposed PILAR* | ✓ | ✓ | | | |

**Large Language Models in AR:** The emergence of LLMs has opened new possibilities for natural language interaction in AR. Recently, several studies have incorporated LLMs into AR applications to enhance functionality. For instance, Lee et al. [19] used an LLM in GazePointAR to resolve pronoun ambiguity in AR voice commands, and Janaka et al. [17] built a wearable assistant (e.g., TOM) that answers user queries about daily activities. While Srinidhi et al. [28] integrated LLMs with AR to develop a cognitive assistant application that guides users through tasks (e.g., making coffee). Other systems have explored LLM-based AR interactions, such as in-context content creation [8] and multimodal object recognition [13]. However, prior works have used LLMs for AR system development; their role as an explanation method remains underexplored. Leveraging LLMs for post-hoc explanations can significantly improve the interpretability, transparency, and personalization of AI decisions in AR (e.g., recipe recommendation). Studies show that such explanations enhance user satisfaction and persuasiveness, particularly when users are unfamiliar with the AI-based recommendation. Motivated by this gap, *PILAR* employs a pre-trained LLM to generate context-aware, personalized explanations in AR, unifying XAI and LLMs for more transparent and user-centric experiences.

## 3 DESIGN CONSIDERATIONS FOR PILAR FRAMEWORK

To generate personalized and human-centric explanations in PILAR, we consider a range of design choices to enhance user interaction and system performance. These details are explained as follows.

### 3.1 Dimensions of Explainability for AR Systems

Explainability goals in AR systems align well with all categories of informativeness in standard XAI [5], such as trust, understandability, persuasiveness, satisfaction, effectiveness, efficiency, and transparency. These dimensions guide the development of intuitive, meaningful, and contextually relevant explanations in intelligent AR applications. *To the best of our knowledge, PILAR is the first framework to adopt these XAI goals within AR, enabling dynamic human-centric explanations tailored for everyday use.* The details of these XAI goals are described below.

**Trust:** Building user trust involves offering consistent, understandable, and privacy-conscious explanations [32], such as clarifying furniture placement choices in AR-based interior design [14].
**Understandability:** Non-experts must grasp how AI-driven elements work in AR, allowing them to address errors, understand data usage, and interact confidently with the system [7, 27].
**Persuasiveness:** Effective explanations use engaging visual and narrative techniques (e.g., storytelling) to align AI suggestions with user goals and preferences, enhancing decision-making [13, 17].
**Satisfaction:** Satisfaction highlights how effectively the AI system meets user expectations, needs, and preferences while providing clear explanations, even for non-expert users (e.g., personalized product suggestions in AR shopping [35]).

**Effectiveness:** Efficient XAI in AR reduces user effort by providing clear, low-cognitive-load explanations, even for those with limited AI literacy [32] (e.g, AR navigation systems can adapt real-time routes, saving time and enhancing usability for non-expert users)

**Transparency:** Transparency in AR ensures users understand AI decisions through clear, jargon-free explanations [4, 19]. It also informs users how their data is collected and used, addressing privacy concerns (e.g., AR navigation apps explain route choices and data sources, allowing user customization and building trust) [32].

### 3.2 Problem Space for Effective XAI-based AR systems

Integrating LLMs into AR systems, especially for crafting an XAI experience, necessitates a nuanced approach considering various problem spaces: *when*, *what*, *how*, *who*, and *where* dimensions. Prior work (e.g., XAIR [32]) covers the first three dimensions without the integration of LLM, while PILAR extends the framework by introducing these two missing dimensions (e.g., *Who* and *Where*) via integrating LLMs and demonstrating how LLM-based explanations can effectively address all of these five dimensions, enabling a more comprehensive and adaptive XAI experience in AR.

**When (Timing of Interaction):** Timely information delivery is essential for effective AR experiences [32]. LLMs can assess user engagement and context to identify optimal moments for interaction. By accounting for users' current activities and cognitive load, LLMs can deliver step-by-step guidance during tasks (e.g., cooking) or minimize cues during cognitively demanding scenarios (e.g., navigation), enhancing personalization while reducing cognitive overload.

**What (Content Relevance):** Effective AR experiences depend on delivering content that is contextually relevant and task-specific [32]. LLMs can generate context-specific, human-centric information (e.g., recipe steps or ingredient tips) for non-expert users. They also personalize content based on user history and preferences (e.g., recipe recommendations), improving efficiency and satisfaction.

**How (Mode of Communication):** Effective XAI in AR requires clear, user-friendly communication [28]. LLMs can generate concise, jargon-free explanations tailored to user profiles, enhancing comprehension, trust, and user engagement. The LLM adjusts its language complexity and tone [28] to match the user (simpler for novices, more technical for experts), ensuring each explanation is concise, jargon-free, and relatable.

**Who (Personalization):** Personalization is key to creating meaningful AR experiences [32]. LLMs can tailor content by analyzing users' past interactions, preferences, and communication styles [] (e.g., LLMs can recommend personalized recipes based on user dietary preferences, health goals, and past interactions), making the experience more engaging and effective.

**Where (Contextual Awareness):** Contextual awareness in AR involves the system's understanding and adapting to the user's physical location and environmental circumstances [32]. LLMs enhance this by providing location-specific information (e.g., offering historical insights or nearby recommendations during navigation) or incorporate spatial context (referencing nearby objects or the user's environment) to enhance relevance, making AR experiences helpful and engaging in everyday life [13, 19].

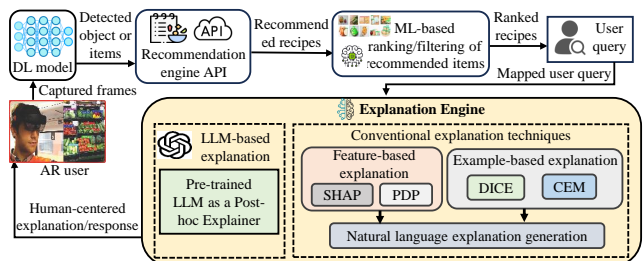

Figure 2: An overview of the PILAR framework.

### 3.3 Mapping Explanation Strategies to User Queries

Mapping user queries to appropriate explanation techniques is critical for effective XAI in AR applications [32], enabling tailored responses to questions such as *"why"*, *"why not"*, and *"what if"* [21, 32]. In HCI, prior work has explored *"why"* questions to feature-based explanations (e.g., SHAP) and *"how to be that"* questions to example-based methods (e.g., CEM). However, this mapping remains underexplored in AR contexts. Template-based approaches (e.g., SHAP, CEM, etc.,) are often rigid, requiring pre-defined methods for each query type. In contrast, using pre-trained LLMs as post-hoc explanation technique, denoted as *LLMEX* offers a unified, flexible approach. LLMEX enables dynamic, personalized, and context-aware explanations across diverse user queries, overcoming the limitations of fixed templates and making it particularly well-suited for AR applications. Details on how user queries map to explanation strategies are provided below.

**Why Not and What-If:** Example-based techniques (e.g., CEM, DICE) or LLMEX can effectively explain why certain recommendations were excluded and how alternatives could be generated. For instance, when a user asks, *"Why wasn't a vegetarian suggested?"*, it can highlight relevant features (e.g., presence of meat) to clarify the decision. Similarly, for *"What if"* queries (e.g, *"What if I want a gluten-free alternative?"*), the system can adapt recommendations accordingly, enabling users to explore personalized alternatives.

**How to be that:** Users interested in emulating a specific recommendation can benefit from example-based explanations. Counterfactual methods (e.g., DICE and CEM) and LLMEX can generate actionable insights by showing how users can modify their behavior or input to achieve a desired recommendation. For instance, a user wonders, *"How can I get recommendations for healthier options?"* The AR interface may suggest modifications like reducing sugar or opting for whole grains, demonstrating the user-specific adjustments needed for preferred recommendations.

**Why:** Feature-based methods (e.g., SHAP) or LLMEX can help users understand why a recommendation was made by highlighting influential features. For instance, a user asks, *e.g., "Why is this recipe recommended?"*, the AR interface can highlight specific ingredients or preparation steps that contributed to the recommendation, offering transparency into the decision-making process.

## 4 IMPLEMENTATION OF PILAR FRAMEWORK

This section presents the implementation of the *PILAR* framework for personalized recipe recommendations and explanation tasks in AR. The system integrates three key modules: (1) ingredient detection, (2) recipe recommendation, and (3) a dual-mode explanation interface (e.g., Template-based and LLMEX), as shown in Figure 2.

### 4.1 System Architecture

**Ingredient Detection:** The system utilizes the YOLOv8-small object detection model [29], which runs directly on a Samsung Galaxy S10 smartphone (Snapdragon 855, 8 GB RAM) for real-time ingredient detection. YOLOv8 is chosen for its balance of speed and accuracy, as it has been shown to achieve real-time performance on edge devices for AR tasks [22]. We train YOLOv8 on a custom dataset of 36 common kitchen ingredients, achieving an F1-score of approximately 90%. At runtime, the camera feed is processed at 15–20 frames per second, with ingredients detected and localized in real time using bounding boxes, as shown in Figure 3a. The detected ingredients are then passed to the recipe recommendation engine.

**Recipe Recommendation:** The recommendation module calls the Edamam Recipe Search API [1] with two inputs: (i) the list of detected ingredients and (ii) the user's dietary profile (e.g., diet type such as vegetarian, health goals like high fiber, or allergy restrictions). The Edamam API applies AI-driven filtering to return personalized recipe suggestions that align with the detected ingredients and user preferences. At runtime, the API responds within 1–2 seconds, providing a ranked list of recipe options sorted by relevance and healthiness. The top-ranked recipe or a small selection of top candidates is then displayed within the AR scene as a floating recipe card anchored near the detected ingredients, as shown in Figure 3b.

**Explanation Interface:** Once a recipe is recommended, users can request explanations via natural language through voice input (processed with on-device speech-to-text) or text input. User queries (e.g., *"Why was this recipe recommended?"*) are combined with contextual information, including detected ingredients, recipe metadata, and user preferences, into a single context object. Based on the selected explanation mode, this context is routed to either a template-based or LLM-based explanation engine, which generates a textual explanation displayed within the AR interface alongside relevant scene objects in real time. This interface allows users to receive justifications directly within the AR environment, aligned with the objects they are interacting with. For instance, if a user asks *"Why not a beef recipe?"*, the system responds with an explanation referencing the user's vegetarian preference, as shown in Figure 3. and may sound formulaic or generic compared to human-like reasoning.

*Template-Based Explanation Engine:* The template-based explanation engine generates an explanation regarding the recommended recipe by populating predefined sentence structures with relevant data. Following prior work [32], we employ two types of explanation techniques: feature-based and example-based. For feature-based explanations, we use two post-hoc methods: SHAP and PDP. SHAP assigns values to features based on their contribution to a model's output using cooperative game theory, helping explain recipe recommendations by highlighting key ingredients or dietary preferences. While PDP illustrates how changes in a single feature (e.g., calories) affect predictions, providing intuitive visualizations of feature impact. For example-based explanations, we employ two post-hoc methods: DICE and CEM. DICE creates alternative input scenarios to show how modifying ingredients would change recommendation outcomes, while CEM identifies features that must be present or absent (e.g., sugar content, allergens) to justify inclusion or exclusion, offering contrastive reasoning. These explanations are generated using structured inputs, including detected ingredients, user profiles, and selected recipes, as shown in Figure 4 of the template-based explanation prompt. For instance, in recipe recommendation tasks, feature-based explanation methods explain Edamam API results by addressing queries (i.e., *"Why was this recipe recommended?"*), clarifying how inputs (e.g., ingredients, preferences) influence outcomes. Likewise, for a user query (e.g., *"How can I modify this recipe to suit my dietary restrictions better?"*), DICE suggests ingredient substitutions (e.g., low-sugar options) aligned with user goals. While these methods enhance transparency and are computationally efficient, they rely on predefined content and may appear formulaic or less natural compared to human-centric (LLMEX).

*LLM-Based Explanation Engine:* To generate real-time explanations for recommended recipes, we utilize a pre-trained GPT-4o mini LLM [3] and apply an in-context learning (ICL) strategy, following prior work [18]. ICL enables the LLM to adapt to new tasks (e.g., recipe explanation) without retraining, by leveraging a few task-specific examples within the prompt. This enables the GPT-4o mini (8B parameters) to generate context-aware, human-centric, and personalized explanations tailored to the needs of AR users. The explanations are generated using prompt engineering techniques, with a temperature of 0.2 and a maximum token limit of 1000. ICL is computationally efficient, making it suitable for real-time AR applications on edge devices, where latency and processing power are limited. As shown in Figure 5, LLM-based explanations incorporate broader context such as dietary restrictions, preferences, past choices, and time of day, unlike template-based methods, which often produce generic outputs (e.g., referencing ingredient matches or health tags). For instance, when the user asks *"Why is this recipe recommended?"*, the LLMEX rephrases the query and includes a contextual summary: *"The user has ingredients <[A, B, . . . ]>, the recommended recipe is <[Recipe Name]>, which includes <[A, B,. . . ]> and fits the user's <[dietary profile]>."* By understanding the user's holistic needs, LLMs offer richer, more personalized, and adaptive explanations, fostering greater user satisfaction and trust. While this extensive contextual capability makes direct comparisons with traditional methods challenging, our goal is to assess how effectively LLMs can utilize the same input features to generate enhanced, user-tailored explanations in AR applications.

## 5 EXPERIMENTAL DESIGN FOR PILAR EVALUATION

This section presents the implementation of the PILAR prototype and the user study design to evaluate its effectiveness in AR-based personalized recipe recommendation and explanation tasks.

### 5.1 System Implementation

We prototype our AR recipe recommendation application on an Android smartphone (e.g., Galaxy S10) with hardware comparable to AR HMDs (e.g., Meta Quest Pro). We developed the app in Unity 2022.3 to leverage smartphone AR features and released it as an open-source mobile application. It performs on-device ingredient recognition, recipe recommendation, and explanation generation, with an intuitive interface built on Unity's Canvas for seamless user interactions. For ingredient detection, the system utilizes OpenCV [2] with a custom YOLOv8 model, which runs on-device, to draw bounding boxes and labels (TextMeshPro) around detected items. Detected ingredients are sent to the Edamam API [1] for recipe recommendation filtered by the user's diet and calorie preferences. To explain recommended recipes, we implemented two explanation modes: (i) an LLM-based approach via real-time GPT-4o mini API calls [3] for personalized explanations, and (ii) a template-based approach on a backend server.

### 5.2 Participants

Our user study involves 16 volunteer participants who received no reward (12 males and 4 females) aged from 20 to 35 years (mean age 23.6, standard deviation (SD) 4.53). The participants were Asian (65.86%), African-American/Black (18.85%), and Hispanic (12.29%). Four of the sixteen participants said they had previously used smartphone-based AR devices.

### 5.3 Study Task and Procedure

Our user study was approved by the Institutional Review Board (IRB) at the authors' university. It was conducted in a controlled indoor lab with uniform lighting and minimal noise, using a Samsung Galaxy S10 smartphone for AR interactions, with researcher guidance provided to ensure consistency. Participants provided informed consent and completed a brief demographic questionnaire (age, gender, etc.,) before commencing. They first received an introduction to the *PILAR* AR recipe recommendation system, which included a brief tutorial on the AR interface and its object-detection feature, followed by a short practice session to familiarize themselves with viewing detected ingredients, using the interface, and receiving diet-aware recipe suggestions. On average, the entire study session lasted about 20 minutes.

**Introduction and Setup:** We designed a within-subjects user study to evaluate the effectiveness and user experience of our *PILAR* system, which supports two explanation modes for recipe recommendations: (1) LLM-generated, human-centric personalized explanations, and (2) Conventional, template-based explanations. The order of conditions was counterbalanced across participants to prevent order effects. The study aimed to compare the effectiveness of different explanation methods in supporting user trust, understanding, and interaction in AR-based recipe recommendation and healthy diet scenarios. Participants were introduced to the PILAR system through a brief tutorial covering the AR interface, object detection,

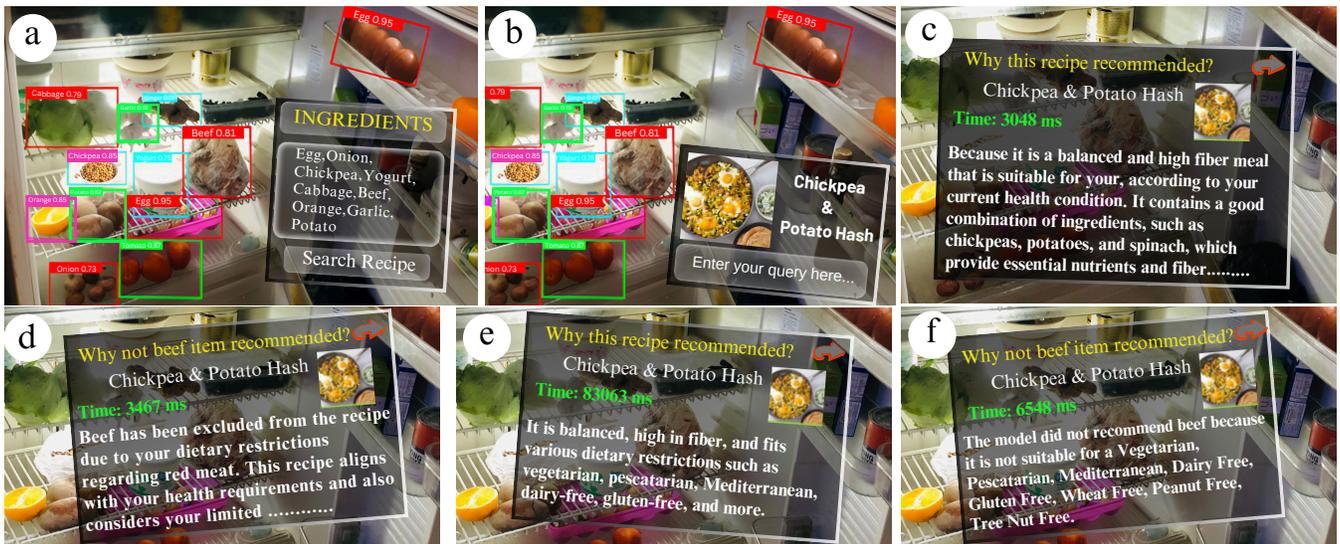

Figure 3: A user opening the fridge to make dinner via our developed PILAR framework UI for reliable recipe recommendation tasks **(a)** The object is detected in real-time, **(b)** The recipe is recommended based on the detected ingredient, and the recipe is ranked based on the user preference (e.g., calorie), **(c)** Explanation generation using LLM interface by asking queries (*"Why this recipe recommended ?"*), **(d)** Explanation generation using LLM interface by asking queries (*"Why not beef item recommended?"*), **(e)** Explanation generation using the traditional explanation interface asking the same question in (c), and **(f)** Explanation generation using the traditional explanation interface asking the same question in (d).

> *Explain briefly why the recipe <recipe name> was recommended to a user with a <dietary preference> diet and a health goal of <health goal>, considering available ingredients: <ingredient list>. Provide the key reason(s) based on matching dietary and nutritional features.*

Figure 4: Prompt for Template-based explanation generation.

> *Please provide a concise, logical, and personalized explanation tailored to a user with a <dietary preference> diet and a health goal of <health goal>. The user currently has these ingredients: <ingredient list>. The suggested recipe is <recipe name>, labeled as <recipe tags>. Clearly address the user's query: <user question>.*

Figure 5: Prompt for LLM-based explanation generation.

and system interaction, followed by a short practice session to explore key functions such as viewing detected ingredients, navigating the interface, and receiving diet-aware recipe suggestions. Using a within-subjects design, each participant experienced both LLM and template-based explanations under consistent conditions. The study targeted non-expert users with low AI literacy to evaluate the system's accessibility and usability for a general audience. Participants received personalized recipe suggestions based on detected ingredients and dietary goals in each explanation setting.

**Tasks:** Participants interacted with the PILAR system in both LLM and conventional settings for recipe recommendation tasks. Across both conditions, they primarily focused on prompts related to dietary restrictions (e.g., *"Why was this recipe recommended?"*), health benefits (e.g., *"What are the health benefits of this recipe?"*), etc., The details of the tasks for both settings are described below.

*Task 1:* Participants began by entering their dietary preferences (e.g., low-carb, vegan, or gluten-free) and health goals into the PILAR system via the interface. Based on their preferences, the system recommended several recipes. In both explanation settings conditions, participants asked, *"Why was this recipe recommended?"* and received explanations regarding the recommended recipes. In the LLM condition, participants received highly personalized, context-aware explanations generated by the LLM model. In contrast, the template condition provided structured rationales based on key features used for recommending recipes.

*Task 2:* Next, participants selected a recipe not recommended by the PILAR system and asked, *"Why wasn't this recipe recommended?"* in both setting. In the LLM setting, the system generated a detailed explanation highlighting why the specific recipe did not meet the user's preferences and goals, providing human-like reasoning for exclusion. In the conventional setting, the system generated a counterfactual explanation using DICE, indicating which specific features would need to change (e.g., fewer carbohydrates) to include the recipe in the recommendations.

*Task 3:* In this task, participants were asked, *"How can I modify this recipe to better suit my dietary restrictions or preferences?"*. In the LLM condition, the system analyzed the recipe and suggested personalized modifications, such as ingredient substitutions or preparation changes, aligned with the user's dietary goals. In contrast, the conventional setting employed example-based methods (e.g., CEM) to present recipes that were more similar to the user's preferences.

*Task 4:* In this task, participants asked free-form questions about the recipe recommendations, such as *"What are the health benefits of this recipe?"* or *"How does this align with my long-term diet goals?"*. In the LLM condition, the system generated comprehensive explanations combining reasoning and personalized insights tailored to each participant's profile. In contrast, the conventional setting relied on predefined techniques to provide more limited responses.

**Post-task Interview Phase:** In the post-task interview phase, we conducted an in-depth, qualitative interview with participants. We asked to complete the system usability scale (SUS) questionnaire. Furthermore, we interviewed and collected verbal feedback from participants for about 3-4 minutes about their reflections on the study, particularly their experience on a 5-point Likert-scale for several dimensions of explainability, discussed in Section 3.1: understandability (*e.g., Clearly and easily to comprehend the provided explanations.*), trustworthiness (*e.g., The provided explanations inspire confidence in the recipe recommendations.*), effectiveness (*e.g., The explanation was useful and in helping you make informed decisions or modify recipes.*), transparency (*e.g., You understood how the system made its decisions and recommendations.*), satisfaction (*e.g., Overall satisfaction with the explanations provided in each condition*), etc., Participants also provided comparative feedback between the two conditions, commenting on which system they preferred and why. This feedback will be crucial in assessing whether the LLM-based explanations provide a more human-centric and user-friendly experience than the template-based explanation, such as feature-based and example-based explanations used in conventional settings in the PILAR framework. Furthermore, participants completed a survey evaluating their interactions after completing all tasks in a given condition (LLM-based and conventional), adapted from the Human-AI Language-based Interaction Evaluation (HALIE) framework proposed by Lee et al [20]. In this survey, participants rated their agreement with the following statements (among others) on a 5-point Likert scale.

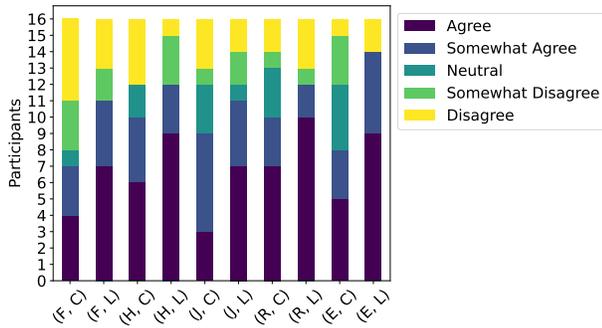

Figure 6: Likert-scale results of the HALIE survey between LLM-based explanations generation condition denoted as **(L)** vs. conventional explanation generation condition denoted as **(C)** in the recipe recommendations tasks in AR.

Table 2: User ratings mean and standard deviation (SD) of the explanation variants PILAR with LLM-based explanation technique denoted as **LLMEX** vs. PILAR with conventional explanation method denoted as **CONV**, rated on a 5-point Likert scale along with P-values of the Wilcoxon signed-rank tests of these two methods.

| Metrics | Mean (SD) | | LLMEX vs CONV |
|---|---|---|---|
| | LLMEX | Baseline | P-value |
| *Understandability* | 4.62 (0.48) | 3.05 (0.78) | 0.00064 |
| *Effectiveness* | 4.56 (0.49) | 2.92 (0.74) | 0.00096 |
| *Transparency* | 4.16 (0.63) | 3.31 (0.91) | 0.00338 |
| *Trust* | 4.48 (0.5) | 2.25 (1.19) | 0.00089 |
| *Efficiency* | 3.95 (0.89) | 2.11 (0.99) | 0.00262 |
| *Persuasiveness* | 4.08 (0.84) | 2.38 (1.31) | 0.00467 |
| *Satisfaction* | 4.43 (0.496) | 2.31 (1.44) | 0.00208 |
| *Overall Satisfaction* | 4.26 (0.74) | 2.18 (1.28) | 0.00319 |

## 6 RESULTS

This section compares the developed PILAR LLM-based explanation UI with the PILAR traditional explanation UI during recipe recommendation tasks. We begin by analyzing our experimental data to study users' reactions when encountering tasks discussed in Section 5.3, using PILAR in AR experiences. Then, we analyzed the post-task user study experimental data after encountering tasks in the AR experience. The details are described in the following.

### 6.1 Time

Participants using the LLM-based explanation setting in the PILAR UI require significantly less time (mean (M) = 163 seconds (s), standard deviation (SD)=36s) to complete all tasks than the participants using the conventional explanation technique setting (M = 310.2s, SD = 51s). To test the significance of these differences in means, we conduct a Shapiro-Wilk test, after which we determine that the obtained data are not normally distributed. Therefore, we conduct the Wilcoxon signed rank tests on these data to quantify the difference between the two conditions' task completion time. We find a significant difference (p-values < 0.05) while comparing these two conditions' task completion time ($W = 2.1$, $p = .00044$ ). This translates to roughly 40% in task completion time on average for PILAR UI conventional explanation setting users compared to PILAR UI LLM-based explanation setting.

### 6.2 Quality evaluation of the explanations during the recipe recommendation

To evaluate the quality of personalized explanation generation for the PILAR framework developed UI with LLM-based and conventional methods for recipe recommendation tasks, we compute the mean ratings for all investigated dimensions (e.g., understandability, trustworthiness, effectiveness, etc.) for each method. The mean and standard deviation of the average ratings of each of these dimensions are reported in Table 2. In all cases, the average ratings are higher for the LLM-based explanation method than the conventional one. Furthermore, we perform the Wilcoxon signed-rank test to measure whether the ratings of the LLM-generated explanations are significantly higher. Before that, we analyze the participants' average ratings normality test to examine the score distribution using the Shapiro-Wilk test and ensure that the sample data is not drawn from a normally distributed population. Table 2 shows the measured P-values of the Wilcoxon test. The results indicate that the obtained ratings are significantly higher for the LLM-generated explanations method than the conventional one (p-values < 0.05). Thus, these results highlight that the quality of the LLM-generated explanations has been higher than the conventional ones. The explanations seemed more understandable because LLMs generate more personalized and context-aware explanations, seamlessly adapting to the user's preferences and inquiry style. In contrast, conventional explanation methods (e.g., SHAP, LIME, etc.,) are limited to feature-based or example-based explanations, which follow a strict pattern and are less tailored to individual users. These methods lack the depth of narrative or reasoning that LLMs provide. That is why the overall satisfaction of the recommender system has been perceived as higher, given the LLM-based explanation. Furthermore, the LLM-based explanation emphasizes user preferences, which explains the higher effectiveness of this recommendation in terms of how well it helps the user evaluate the suggested recipe. Indeed, LLMs can offer real-time reasoning across various user queries, enhancing the perception of trustworthiness and effectiveness during the recipe recommendation tasks. Overall, the results indicate that LLM-generate personalized explanations offer a more holistic and user-centered approach, making them more favorable for recipe recommendations in AR, where users seek trustworthy, and efficient interactions.

### 6.3 HALIE Survey and System Usability

We also provide additional user study results demonstrating how well the PILAR UI LLM-based explanation generation is better than the conventional method-based explanation for the recipe recommendation tasks. For this, we asked users to rate their experiences using the HALIE survey questionnaire on a 1 to 5 Likert scale during the user study. To evaluate the rated users' experiences, we compute the mean and standard deviation of the Likert score for each statement (e.g., fluency (F), helpfulness (H), enjoyment (J), responsiveness (R), and ease (E)) for each user from the HALIE survey. The user feedback results are provided in Figure 6. We observe that most users preferred and approximately 85% agreed with all statements of the condition with LLM-based explanations than the condition with conventional-method-based explanations in the recipe recommendation tasks.

We also analyze the participants' Likert score normality test to examine the score distribution using the Shapiro-Wilk test and ensure that the sample data is drawn from a normally distributed population. Therefore, we conduct the t-test to quantify the difference between these two conditions' Likert scores. We find a significant difference between conditions with LLM-based explanations and conditions with conventional-method-based explanations on all factors except responsiveness (R) factors in the HAILE survey (p-values < 0.05). The p-values for these two conditions are all factors ($F = 0.007$, $H = 0.014$, $J = 0.027$, $R = 0.053$, and $E = 0.0014$). Thus, these results show that, compared to the conventional explanation technique, recipe recommendation using LLM-based explanation could effectively improve the intelligibility, transparency, and trustworthiness of AR systems for end-users.

To quantify the utility and usability of the PILAR UI LLM-based explanation condition with the conventional explanation method, we used the System Usability Scale (SUS) [6], since it was initially designed for evaluating new tools or systems. The questionnaire used a five-point Likert scale to register the feedback of the user testing participants. The academic literature indicates that a score of 68 is deemed an average SUS score for system usability studies [6]. A score greater than 68 shows high system usability. A score less than 68 indicates that there are problems with the usability aspects of the system under test. The descriptive statistics (e.g., mean, standard deviation, and 95% confidence interval (CI)) of conditions with LLM-based explanations and conditions with conventional-method-based explanations measured SUS final scores are shown in Table 3. The results show that the PILAR UI LLM-based explanation condition receives a mean high SUS score of 82.5, denoted a grade A (excellent), indicating excellent usability of the system. The conventional explanation method receives a mean average SUS score of 71.5, denoted a grade B (Good) in terms of usability. Furthermore, we also analyze the participants' SUS score's normality test to examine the score distribution using the Shapiro-Wilk test and ensure that the sample data is drawn from a normally distributed population. Therefore, we conduct a t-test and find a significant difference between the conditions with LLM-based explanations and conditions with conventional-method-based explanations' SUS scores ($t = 5.2$, $p = 0.001$). Thus, these results show that participants have excellent usability of the system when using the PILAR with LLM-based explanation conditions rather than PILAR with conventional-method-based explanation conditions.

Table 3: Descriptive statistics SUS final scores between LLM and conventional-method based explanations.

| Condition | Mean (SD) | 95% CI |
|---|---|---|
| LLM | 82.4 (6.89) | [79.2, 85.4] |
| Conventional | 71.5 (8.05) | [65.6, 74.2] |

## 7 DISCUSSION

This section briefly discusses the results obtained using the proposed PILAR framework, which leverages pre-trained LLM to generate personalized, human-centric explanations regarding recipe recommendation tasks. By delivering real-time, context-aware reasoning, PILAR addresses key limitations of traditional methods, offering intuitive, relevant, and preference-aligned explanations. Implemented in a real-world AR setting, PILAR's LLM-based explanation interface significantly outperformed the traditional approach in terms of user satisfaction, trust, and system usability. Participants appreciated the holistic nature of the LLM-based reasoning. As **P2** noted, *"PILAR gave me everything I needed at once (e.g., why this recipe is good for my diet), which was clear and personalized."* In contrast, **P4** noted, *"I found the traditional method confusing and I had to jump between explanations to piece everything together."* Overall, feedback from usability assessments and the HALIE survey confirmed that most users preferred PILAR's LLM-based explanations over traditional methods, highlighting their effectiveness in enhancing the AR recipe recommendation experience (see Figure 6 and Table 2). Since no prior work has applied pre-trained LLMs to generate high-quality, context-aware, human-centric explanations in AR environments [13, 16, 17, 19, 25, 32], our results are not directly comparable to existing literature. However, we evaluated user preferences and real-time interactions to assess the effectiveness of our proposed PILAR framework. Prior research has primarily relied on static, template-based explanation techniques, which lack adaptability across key XAI dimensions [24, 32]. These approaches often fail to meet users' evolving needs in dynamic AR environments. In contrast, PILAR's LLM-based approach delivers flexible, personalized, and context-aware explanations that adapt to real-time user interactions, making it well-suited for dynamic AR scenarios (see Table 2). Participants consistently found LLM explanations more intuitive and aligned with their preferences than traditional methods. For instance, **P5** noted, *"PILAR made my recipe recommendations feel personalized. It felt like I had a personal nutritionist guiding me."*. While the same participant noted that traditional explanations felt fragmented, each method addressed a different aspect. These results indicate that the LLM-based explanation method provided a more cohesive, engaging, and informative user experience (see Figure 6 and Table 3).

*Applicability of the PILAR Framework to Other AR Applications:* Although implemented for personalized recipe recommendations, the PILAR framework is adaptable to a variety of AR scenarios, such as furniture assembly assistance and fitness coaching. By delivering personalized, human-centered explanations, PILAR can enhance user engagement and interaction across diverse AR contexts. For instance, it could offer step-by-step, skill-aware instructions in furniture assembly or provide real-time, adaptive feedback in fitness coaching based on user performance and preferences. While these domains were not explored in this study, qualitative feedback suggests strong potential for PILAR's broader applicability in delivering context-aware, intuitive AR experiences.

## 8 LIMITATIONS AND FUTURE WORK

Our proposed PILAR framework has a few limitations. First, our user study involved a small, demographically limited participant pool focused solely on recipe recommendation tasks, with imbalanced gender representation and no analysis of factors such as demographic background, AI/AR experience, or AI literacy [32]. Although we collected demographic data (e.g., age, gender, ethnicity), their influence on user behavior was not examined. Future work should investigate how individual differences affect engagement, explore broader participant groups, and evaluate PILAR across diverse tasks for more comprehensive insights. Second, our prototype was evaluated exclusively on a smartphone-based AR platform. While PILAR is designed to be hardware-agnostic, its performance on other AR devices (e.g., Meta Quest Pro, Apple Vision Pro, etc.) remains untested. Given the variability in computational power, display quality, and interaction methods across AR devices, subsequent studies should evaluate PILAR's effectiveness and adaptability on these varied platforms. Our study also did not explicitly examine how user trust in AI evolves over extended usage, a critical aspect for sustained user engagement, especially in sensitive application domains such as health or fitness coaching. Finally, real-world deployment raises significant privacy and ethical challenges, including unintended capture of bystander data and secure handling of sensitive user information [32], highlighting the need for advancements in secure on-device computing.

## 9 CONCLUSION

This paper proposed *PILAR*, a novel framework that leverages pre-trained LLMs to generate personalized, human-centric explanations, enabling more intuitive and trustworthy real-time interactions in AI-powered AR systems. Unlike traditional template-based methods that rely on multiple explanation techniques, PILAR offers a unified, context-aware reasoning approach that dynamically adapts to user needs. We implemented PILAR in a smartphone-based AR application for recipe recommendation tasks, an open-source prototype system that integrates computer vision, AI-enabled recipe suggestions, and pre-trained multimodal LLMs for explanation. A user study with 16 participants demonstrated that PILAR significantly outperforms traditional template-based explanation interfaces, with participants completing tasks 40% faster and reporting higher satisfaction. By anchoring AR content with LLM-driven explanations, PILAR paves the way for personalized, trustworthy, and immersive AR experiences in everyday contexts.

## 10 ACKNOWLEDGEMENT

This material is based upon work supported by the Army Research Office (ARO) under award number W911NF-23-1-0401. Any opin-ions, findings, conclusions, or recommendations expressed in this publication are those of the authors and do not necessarily reflect the official policy or position of the Army, Department of Defense, or the U.S. Government.